\documentclass[aps,prb,preprint,superscriptaddress,groupaddress,eqsecnum,floatfix]{revtex4-1}

\usepackage{graphpap}
\usepackage{graphicx}
\usepackage{color}
\usepackage{hyperref}
\usepackage{amsmath}
\usepackage[normalem]{ulem}
\usepackage{wasysym} 
\usepackage{comment}
\usepackage[separate-uncertainty = true, 
]{siunitx}
\graphicspath{{./figures/}}

\usepackage{xcolor}

\usepackage{xcolor}

\usepackage{xcolor}

\usepackage{nomencl}
\makenomenclature

\begin{document}

\title{Aharonov-Bohm oscillations and magnetic focusing in ballistic graphene rings}

\author{Jan Dauber}
\affiliation{JARA-FIT and 2nd Institute of Physics, RWTH Aachen University, 52074 Aachen, Germany}
\affiliation{Peter Gr\"unberg Institute (PGI-9), Forschungszentrum J\"ulich, 52425 J\"ulich, Germany}

\author{Martin Oellers}
\affiliation{JARA-FIT and 2nd Institute of Physics, RWTH Aachen University, 52074 Aachen, Germany}

\author{Florian Venn}
\affiliation{JARA-Institute for Quantum Information at RWTH Aachen University, Aachen, Germany}

\author{Alexander Epping}
\affiliation{JARA-FIT and 2nd Institute of Physics, RWTH Aachen University, 52074 Aachen, Germany}
\affiliation{Peter Gr\"unberg Institute (PGI-9), Forschungszentrum J\"ulich, 52425 J\"ulich, Germany}

\author{Kenji Watanabe}
\affiliation{National Institute for Materials Science, 1-1 Namiki, Tsukuba, Japan}

\author{Takashi Taniguchi}
\affiliation{National Institute for Materials Science, 1-1 Namiki, Tsukuba, Japan}

\author{Fabian Hassler}
\affiliation{JARA-Institute for Quantum Information at RWTH Aachen University, Aachen, Germany}

\author{Christoph Stampfer}\email{Corresponding author: stampfer@physik.rwth-aachen.de}
\affiliation{JARA-FIT and 2nd Institute of Physics, RWTH Aachen University, 52074 Aachen, Germany}
\affiliation{Peter Gr\"unberg Institute (PGI-9), Forschungszentrum J\"ulich, 52425 J\"ulich, Germany}
 
\date{ \today}

\begin{abstract}
We present low-temperature magnetotransport measurements on graphene rings encapsulated in hexagonal boron nitride. We investigate phase-coherent transport 
and show Aharonov-Bohm (AB) oscillations in quasi-ballistic graphene rings with hard confinement.
In particular, we report on the observation of $h/e$, $h/2e$ and $h/3e$ conductance oscillations.
Moreover we show signatures of magnetic focusing effects at small magnetic fields confirming ballistic transport.
 We perform tight binding calculations which allow to reproduce all significant features of our experimental findings and enable a deeper understanding of the underlying physics. Finally, we report on the observation of the AB conductance oscillations in the quantum Hall regime at reasonable high magnetic fields, where we find regions with enhanced AB oscillation visibility with values up to $0.7$\%.
These oscillations are
well explained by taking disorder into account allowing for a coexistence of hard and soft-wall confinement.

\end{abstract}

 \pacs{???}  
 
\maketitle

\newpage

\section{Introduction}
Quantum interference and ballistic transport are corner stones in mesoscopic physics \cite{imr02} giving rise to many interesting effects, such as weak localization corrections, universal conductance fluctuations,\cite{lee85} quantized conductance,\cite{wee88} persistent currents,\cite{ble09} Aharonov-Bohm oscillations \cite{web85} and many others. These effects are known to depend strongly on system size and characteristic length scales of electron transport, such as the elastic mean free path and the phase-coherence length. As these length scales are crucially affected by impurity and defect scattering as well as by electron-phonon and electron-electron interactions, high-purity materials and low temperatures are vital for controlling and utilizing quantum interference phenomena in mesoscopic systems. For experimentally adressing these phenomena two-dimensional (2D) electron systems based on III-V heterostructures have proven as a very favorable platform.

Thanks to the recent advances in process technologies for assembling 2D
materials,\cite{gei13} we nowadays have bulk graphene samples with
elastic mean free paths exceeding values of 25~$\mu m$.\cite{ban16} These
advances not only start closing the gap between graphene and the most
favorable GaAs-based 2D electron-systems but they also open the door to
Dirac fermion-based mesoscopic devices with a hard-wall confinement and
to a unique platform for electron-optics.\cite{agr13,bog17} 
Recently a number of interesting devices and phenomena based on ballistic transport in graphene have been demonstrated, ranging from Fabry-Perot and commensurability oscillations,\cite{you09,han17} 
to magnetic focusing~\cite{tay13} and the Veselago lens~\cite{lee15} to Klein tunneling transistors~\cite{wil14} and even ballistic graphene Josephson junctions.~\cite{bor16}
All these devices rely on high-mobility graphene with large lateral dimensions such that edges
are not limiting transport.

In this work we show that ballistic transport in state-of-the-art graphene can also be maintained when carving out mesoscopic rings, allowing to study electron interference in mesoscopic devices with a truely hard-wall confinement and a widely tunable Fermi wave length. In particular 
 we present low-temperature magneto-transport measurements on graphene rings
 exhibiting ballistic transport, Aharonov-Bohm conductance oscillations and
 magnetic focusing. Aharonov-Bohm (AB) interference~\cite{aha59} experiments
 on ring structures have been proven to be useful for observing and controlling interference patterns as they provide a straightforward way for shifting the phase by a small out-of-plane magnetic field. The AB effect has been studied in detail in rings based on metallic films,\cite{web85} semiconductor heterostructures~\cite{tim87} and more recently also on graphene.\cite{rus08,hue09,hue10,smi12,cab14}
Here it is important to note that
in all previous studies on graphene rings, the samples have been in the diffusive regime, 
which in contrast to ballistic transport (i) leads to a strong suppression of AB oscillations at very
low carrier densities,\cite{rus08} (ii) does not allow to access magnetic
focusing, and (iii) prohibits the probing of the coexistence of hard- and soft-wall confinement.

As graphene only consist of surface atoms, the
sample quality is highly influenced by disorder arising from substrate interaction and edge roughness.\cite{sta09,gal10} By fully encapsulating graphene in hexagonal boron nitride (hBN) the sample quality is improved substantially.\cite{dean10,mayo11,pono11,wang13} Importantly, this improvement also holds for hBN-graphene-hBN structures with lateral dimensions down to around $200$\,nm.\cite{ter16} For smaller structures disorder due to the rough edges of graphene leading to localized states strongly limits high mobility transport.\cite{eng13,bis15,ter16} 
Our work is based on this recent insights and
we report (i) on the fabrication of graphene rings fully encapsulated in hBN, (ii)
 phase-coherent transport with focus on AB oscillations for small magnetic fields and (iii) on the observation of electron guiding and magnetic focusing for intermediate magnetic fields indicating quasi-ballistic transport. Moreover, we perform tight binding calculations of graphene rings with similar geometries and compare the simulation results with the measurements, being able to assign specific skipping orbits to signatures in the measured conductance traces. 
Finally, we discuss highly visible AB oscillations at the onsets of well established quantum Hall plateaus in experiment and theory.

\begin{figure*}[t]\centering 
\includegraphics[draft=false,keepaspectratio=true,clip,%
                   width=0.95\linewidth]%
                   {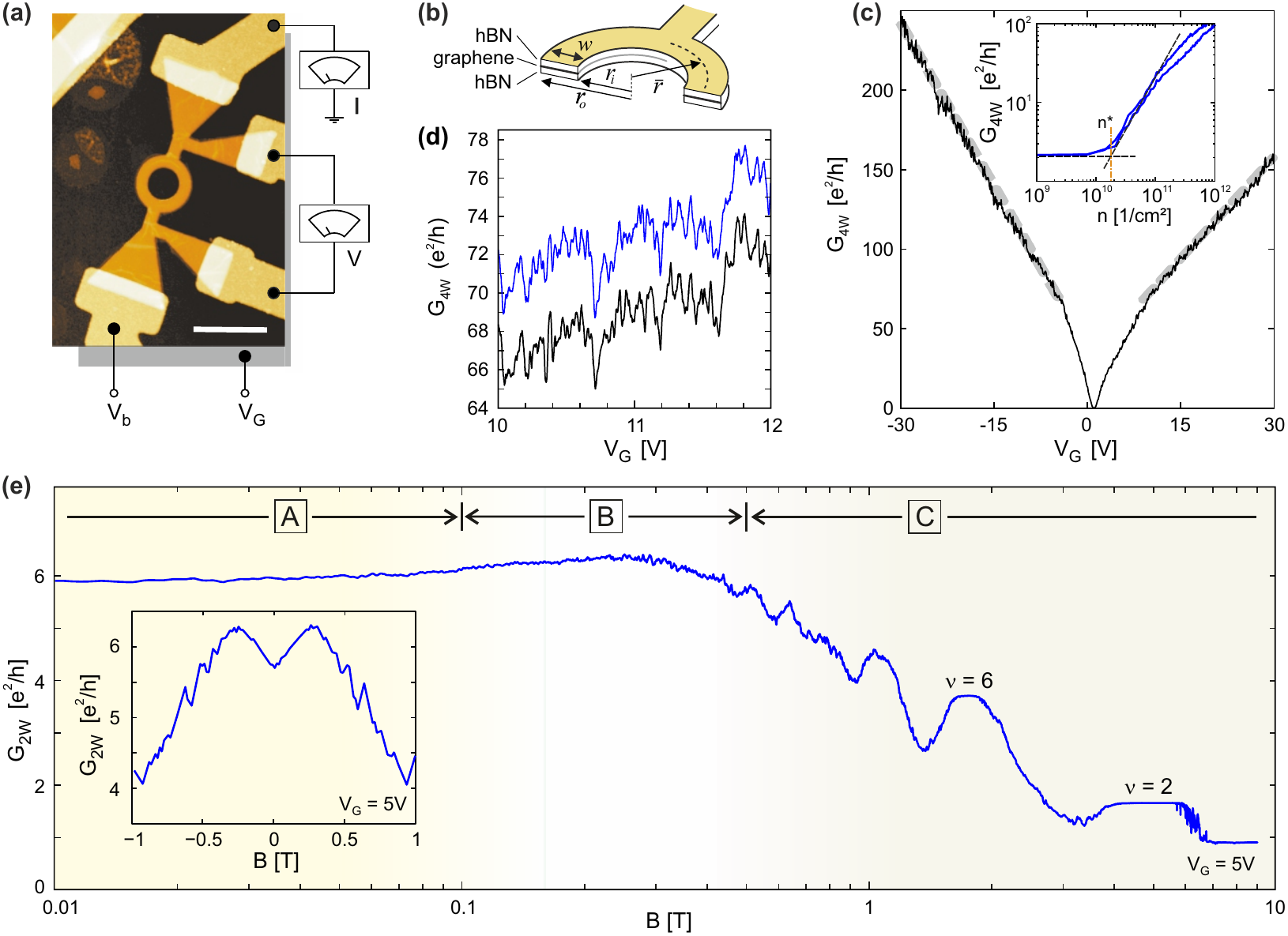}                   
\caption{
(a) Scanning force microscopy image of device \#1 with a schematic of the measurement configuration. The white scale bar is $2~\mu$m. (b) Illustration highlighting the cross-section of the ring device and its dimensions (see text). (c) $G_{4W}$ as a function of $V_G$ at $B=0$\,T. Dashed lines represent fits for extracting the field effect mobility $\mu$. Inset shows log-log plot for the extraction of $n^*$. (d) Repeated measurements of $G_{4W}$ over a smaller range of $V_G$ showing universal conductance fluctuations with good reproducibility. The traces are offset by $4 $e$^2$/h for clarity. (e) $G_{2W}$ as function of $B$ at $V_G=5$\,V. The $B$-field axis is divided into three regimes: low field ($r_C > w$ - label A), intermediate field ($r_C \approx w$ - label B) and high field range ($r_C < w/2$ - label C).  The inset shows $G_{2W}$ for $|B|\leq 1$\,T. All measurements are taken at $T=36$\,mK.
} 
\label{figure1}
\end{figure*}

\section{Methods}

We present measurements on two samples with different geometry. Unless
stated otherwise data from sample \#1 are shown. The samples have been
fabricated by stacking mechanically exfoliated graphene and hBN with the dry
transfer technique based on van-der-Waals adhesion introduced by Wang et
al.~\cite{wang13} on highly p-doped Si substrates with a 285\,nm thick SiO$_2$
top layer. The hBN-graphene-hBN sandwiches have been structured by reactive ion
etching (SF$_6$/Ar plasma) using an AZ nlof 2020 mask patterned by electron
beam lithography (EBL). Electrical contacts to the etched devices have been
made by a second EBL step, metal evaporation (5\,nm Cr/95\,nm Au) and standard
lift-off process. A scanning force microscopy (SFM) image of the final device
\#1 is shown in Fig.~1(a). From the SFM image we extract an inner and an outer
ring radius of $r_i=405$\,nm and $r_o=755$\,nm, respectively (see also Fig.~1(b)). These numbers results in a mean radius of $\overline{r}=580$\,nm and a width of the ring arms of $w=r_o-r_i=350$\,nm. The second sample has slightly different dimensions exhibiting a mean radius of $\overline{r}=500$\,nm and a width of the ring arms of $w=200$\,nm.  
All measurements are performed in a dilution refrigerator with base
temperature of around $T=36$\,mK and perpendicular magnetic field up to $9$\,T
using standard low-frequency lock-in techniques for simultaneous two (2W) and
four-terminal (4W) measurements. The two-terminal current is measured with a
home build I/V converter and a low noise amplifier in combination with a
standard lock-in amplifier, whereas the four-terminal voltage drop is measured
directly by the lock-in as illustrated in Fig.~1(a). For the data analysis of the magnetic field dependent measurement we mainly concentrate on the two-terminal measurements, as the data quality is better thanks to pre-amplification.

The quantum transport simulation have been performed using a tight-binding approximation on a hexagonal lattice with the Kwant package.\cite{kwant} We have simulated a scaled version of the graphene lattice with a lattice constant that is a factor 10 larger than the experimental situation and our simulation involves in total $7\times 10^5$ lattice sites. 
All results are presented in original units for easier comparison to the experimental findings.
For more details on the simulation methodology, see Appendix 1.

\begin{figure*}[tb]\centering 
\includegraphics[draft=false,keepaspectratio=true,clip,%
                   width=0.95\linewidth]%
                   {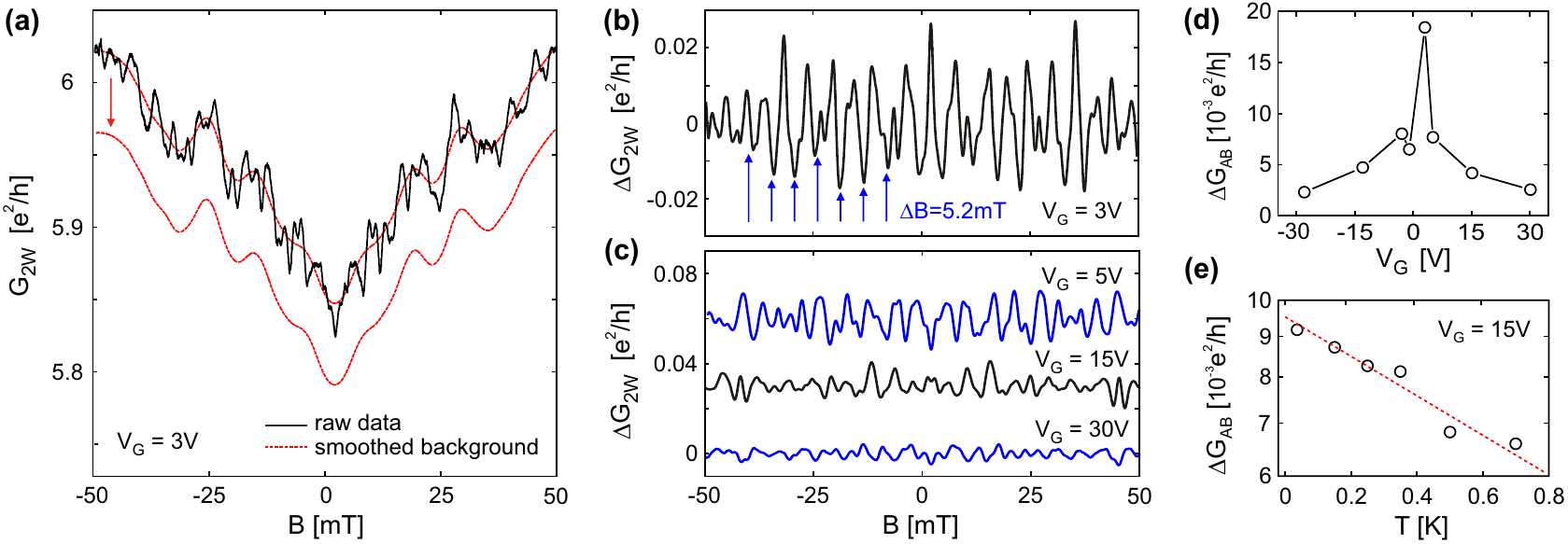}                   
\caption{
(a) Solid black line: $G_{2W}$ as function of $B$ at $V_G=3$\,V. Red dashed lines shows $\left\langle G_{2W} \right\rangle$ used for background subtraction. For clarity the trace is duplicated with an offset (see red arrow). (b) and (c) Processed data with $\Delta G_{2W}$ for various $V_G$. Vertical arrows in panel (b) indicate a periodicity of $\Delta B_{AB}=5.2$~mT. 
In panel (c) the traces are plotted with an offset for clarity.  (d) $\Delta
G_{AB}$ as function of $V_G$. All these measurements are taken at $T=36$\,mK.
(e) $\Delta G_{AB}$ as function of $T$ at $V_G=15$\,V. Red dashed line is a
guide for the eye to highlight the $ \Delta G_{AB} \propto \exp(- c T)$ dependence.
} 
\label{figure2}
\end{figure*}

\section{Results and Discussion}

Figure~1(c) shows the four-terminal back-gate characteristics for zero magnetic
field at base temperature. At high gate voltage $V_G$ we observe a linear
dependence of the four-terminal conductance $G_{4W}$ as function of $V_G$ with
a slope of $\Delta G_{4W}/\Delta V_G$, (see dashed lines). We conservatively
estimate the square conductance $G_{\Box} = e n \mu \approx 4 G_{4W}$ from the
sample geometry~\cite{FNgeometry}. As the charge carrier density $n$ is
directly linked to $V_G$ by $n=\alpha(V_G-V_{CNP})$, we find the field effect
mobility by $\mu \approx 1/(4 e\alpha) (\Delta G_{4W}/\Delta V_G)$. Here,
$\alpha = 6.7 \times 10^{10}$\,cm$^{-2}$V$^{-1}$ is the gate lever arm, which
has been extracted from quantum Hall measurements (see below), $e$ is the
elementary charge and the charge neutrality point (CNP) is around
$V_{CNP}\approx 1$\,V. We obtain field effect mobilities of $\mu_h \approx
100{,}000$\,cm$^2$V$^{-1}$s$^{-1}$ for hole and $\mu_e \approx 60{,}000$\,cm$^2$V$^{-1}$s$^{-1}$ for electron transport.
These mobility values correspond to a mean free path in the range of $l_m \approx 400\,$nm$ - 1.6\,\mu$m (for $V_G=10-30$~V)~\cite{FNmfp} highlighting that the transport can be considered as quasi-ballistic ($w < l_m < l$, where $l$ is the sample length exceeding $\pi \overline{r}$) and that the approximate diffusive description only holds due to diffusive scattering at the rough edges. 
%
With the Einstein relation we estimate the diffusion constant $D =n \mu /(e
\rho)$, where $\rho$ is the density of states of graphene. By following
Ref.~\onlinecite{rus08} we express $\rho$ as function of $n$ and obtain $\rho =
4/(h v_F) \sqrt{\pi n}$, where $v_F=10^6$~m/s is the Fermi velocity and $h$
the Planck constant. For electron transport we thus extract a diffusion
constant of $D = 0.28-0.49$~m$^2$/s (for $V_G=10-30$~V). These values are close
to the value of the diffusion constant only taking into account diffusive
scattering at the edges of the narrow leads and the ring arms ($D= w v_F =
0.35$~m$^2$/s). The high sample quality is moreover reflected in the low
residual charge carrier density $n^\ast$ found to be in the order of
$2\times10^{10}$~cm$^{-2}$, extracted as shown in the inset of Fig.~1(c).\cite{cou14}
By having a closer look at the four-terminal back-gate characteristics we
observe well reproducible universal conductance fluctuations (UCFs) with an
amplitude of up to $2$\,$e^2/h$ (see Fig.~1(d), a close-up of Fig.~1(c)). The
UCFs are a strong evidence for phase-coherent transport. 

In Fig.~1(e) the two-terminal conductance $G_{2W}$ is plotted as function of magnetic field $B$ at $V_G=5$\,V in the near vicinity of the CNP. For small magnetic fields we observe a nearly linear increase of $G_{2W}$ as function of $B$ (see inset in Fig.~1(e)) with $G_{2W}$ reaching a maximum at $B = \pm 0.3$~T. At higher $B$-fields we observe signatures originating from the quantum Hall effect. As the two-terminal conductance $G_{2W}$ contains contributions from the longitudinal $G_{xx}$ and the Hall conductance $G_{xy}$ we observe dips before entering quantum Hall plateaus, which are typical for two-terminal devices, where the sample length $l$ is larger than the width $w$ (sample \#1 $w=350$\,nm to $l \sim$ several $\mu$m).\cite{aba08} Quantum Hall plateaus are well visible for filling factors $\nu=2~\textrm{and}~  6$ (see labels in Fig.~1(d)). The difference in conductance with respect to the theoretically expected plateaus at $2$ and $6~e^2/h$ is due to additional resistances from the setup and the contact resistance of the graphene-metal interface included in the two-terminal measurement. With the known setup resistance $R_S \approx 2.3$\,k$\Omega$, we estimate an overall contact resistance of $R_C \approx 200\,\Omega$ for each contact. 

To systematically discuss the magnetic field dependent conductance measurements we divide the $B$-field range into three regimes based on the ratios of the width of the ring arms $w$ (as well as the mean ring radius $\overline{r}$) and the cyclotron radius $r_C=\hbar \sqrt{\pi n}/e B$ (here $\hbar$ the reduced Planck constant).
First (Sec. III.A) we focus on small $B$-fields such that $w,\overline{r}<\!\!< r_C$. In this regime (see label {\bf A} in Fig.~1(e)) we mainly concentrate on AB oscillations and their dependence on charge carrier density. Second (Sec. III.B) we discuss magnetic focusing and localization effects in the regime, where the cyclotron radius is on the order of $w$ and $\overline{r}$ (i.e. $w,\overline{r} \sim r_C$), see label {\bf B} in Fig.~1(e). Third (Sec. III.C) we focus on quantum interference phenomena in the quantum Hall regime (label {\bf C}), where $r_C$ is significantly smaller than $w$ and $\overline{r}$.  In particular, we require $r_C < w/2$ for entering the regime of quantum Hall edge transport.



\subsection{Aharonov-Bohm oscillations at low $B$-fields}
\begin{figure*}[tb]\centering 
\includegraphics[draft=false,keepaspectratio=true,clip,%
                   width=0.75\linewidth]%
                   {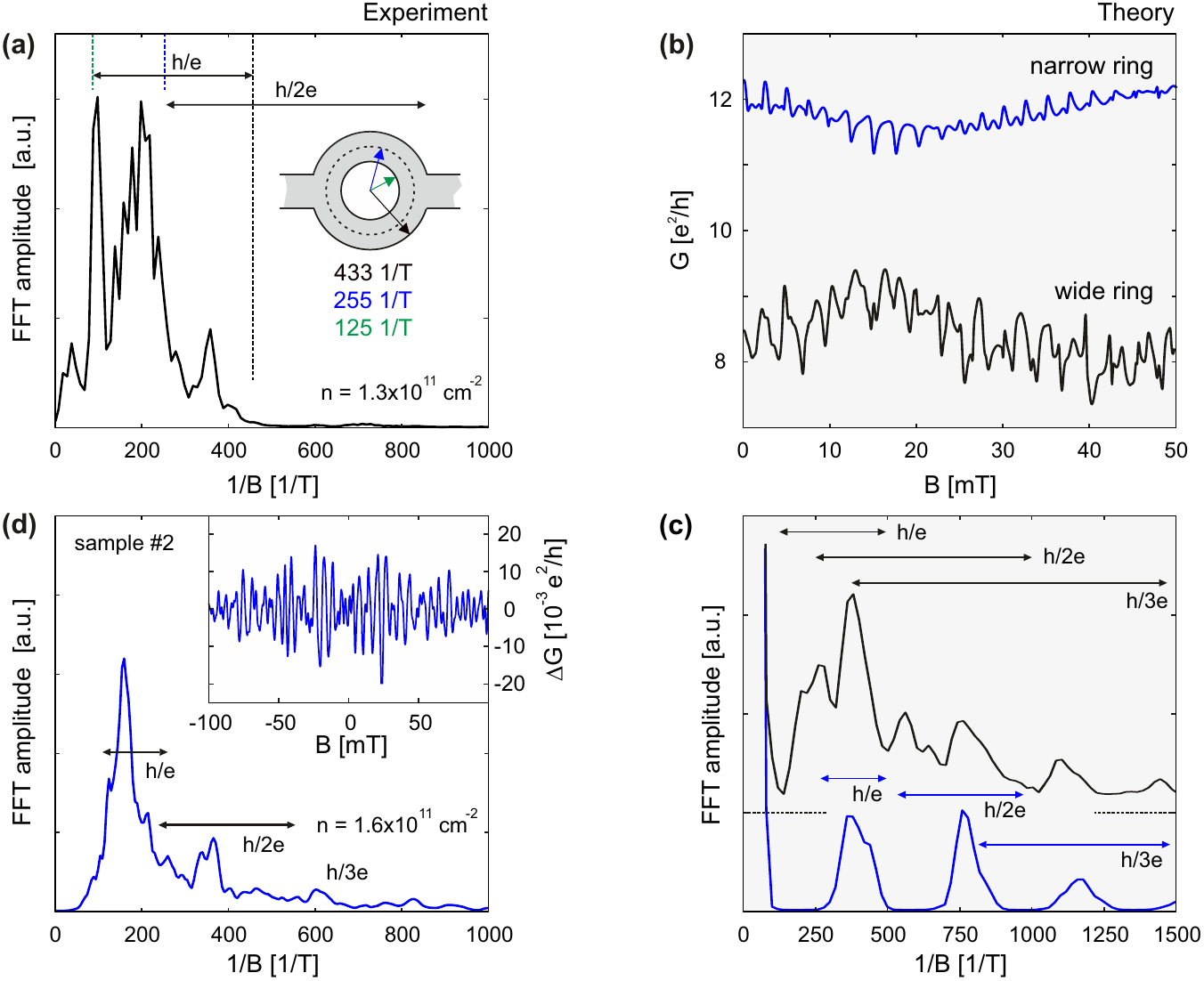}                   
\caption{
(a) Fourier spectrum of the AB oscillations measured at $V_G=3$\,V as shown in
Fig.~2(b). Frequency range of individual AB oscillation modes marked by
arrows. Horizontal lines indicate frequencies for inner, mean and outer radii
as illustrated in the inset. (b) Tight binding simulations of
magnetoconductance of a wide ring ($\overline{r}=600$\,nm, $w=400$\,nm)
similar to sample \#1 and a narrow ring ($\overline{r}=700$\,nm, $w=200$\,nm)
comparable to sample \#2 calculated without edge roughness for $n=1.6 \times
10^{11}$~cm$^{-2}$ (offset by $4e^2/h$). (c) Fourier spectra of the data
presented in panel (b) with indicated AB modes. (d) Fourier spectrum of data
from sample \#2 for similar $n$ as in panel (a). Inset shows background subtracted $\Delta G$.
} 
\label{figure3}
\end{figure*}

Figure 2(a) shows the two-terminal conductance $G_{2W}$ as function of
magnetic field in the small magnetic field range $|B| \leq 50\,$mT at $V_G =
3$\,V. In this regime we observe the presence of AB oscillations as well as
UCFs and an overall increase of the conductance with absolute value of
magnetic field. As the change of conductance due to AB interference is on the
order of a few percent and overlapping with UCFs, additional data
post-processing is needed to extract the periodicity $\Delta B_{AB}$ and the
amplitude of the AB oscillations $\Delta G_{AB}$. We apply a moving average
over a window of $\Delta B = 15~\,$mT, which is sufficiently larger than the
expected periodicity $\Delta B_{AB}=2.3-8$~mT for the $h/e$ mode~\cite{FNabo}
and subtract the averaged data from the raw signal (see red dashed line in
Fig.~2(a)). Fig.~2(b) displays the processed data of the two-terminal
conductance $\Delta G_{2W} = G_{2W} - \left\langle G_{2W}
\right\rangle_{\Delta B}$ versus magnetic field $B$. In this plot horizontal
lines with a spacing of $\Delta B_{AB}=5.2$\,mT indicate the expected
periodicity of the AB oscillations for an ideal ring device with a radius of
$580$\,nm, which corresponds to the mean radius of our fabricated graphene
ring. In general, the conductance oscillations are found to be mainly
equidistant, but also additional modulations and local derivations are
observed. Figure 2(c) depicts similar data, but for different gate voltages
(see labels). Interestingly, for higher gate voltages the AB amplitude
decreases, while the periodicity is preserved. This observation is summarized
in Fig.~2(d), where $\Delta G_{AB}$ is plotted as function of $V_G$. 
Here $\Delta G_{AB}$ is defined as the root mean square amplitude of $\Delta G_{2W}$. Strikingly, the maximum visibility~\cite{FNvisib} is found close to the charge neutrality point at $V_G=3$\,V and is around $0.25\%$ with respect to the overall conductance $G_{2W}$. Although the maximum two-terminal visibility is similar to  earlier experiments on diffusive graphene rings,\cite{rus08,cab14} the carrier density dependency is inverted. For example, in
contrast to Ref.~\onlinecite{rus08} we observe a decreasing amplitude $\Delta G_{AB}$ for increasing gate voltage $V_G$. We explain this behavior by an increasing asymmetric transmission through the two ring arms coming along with an increasing 
 mean free path $l_m$ as function of carrier density. Entering the quasi-ballistic transport regime,  the transmission through each of the two arms depends on the specific microscopic sample shape and disorder potentials. Therefore, the transmission is not symmetric, leading to a reduced visibility of AB oscillation~\cite{vas07}. Also with increasing charge carrier density the Fermi wavelength $\lambda_F = \sqrt{4 \pi/n}$ is decreasing allowing a higher number of wave modes inside the ring, which may also decrease $\Delta G_{AB}$ as it is observed in semiconductor heterostructures.\cite{for88}
In Fig.~2(e), $\Delta G_{AB}$ is plotted as function of temperature $T$ at
$V_G=15$\,V. The decrease of the amplitude $\Delta G_{AB}$ with an increase of
$T$ follows an overall $\Delta G_{AB} \propto \exp{(-c T)}$ dependence. Due to
thermal averaging of the $h/e$ oscillations and changes in the phase-coherence
length $l_{\phi}$, the AB amplitude decays as $\Delta G_{AB} \propto
(E_{Th}/k_B T)^{1/2} \exp(-\pi \overline{r}/l_{\phi})$, where $E_{Th}$ is the
Thouless energy and $k_B$ the Boltzmann constant.\cite{rus08} In a diffusive
system, $E_{Th}$ is given by $E_{Th}=\hbar D / L^2$ with characteristic lenght
scale $L$. Using the diffusion constant determined above $D=0.36$\,m$^2$/s
(for $V_G = 15$~V)and $L=\pi \overline{r}$, we estimate $E_{Th}=71\,\mu$eV,
which corresponds to a critical temperature of $T_c=E_{Th}/k_B=830$\,mK.
Without thermal averaging the phase-coherence length is primarily limited by
electron-electron scattering following a temperature dependence given by
$l_{\phi} \propto T^{-1}$ as also reported for ballistic one-channel rings in
GaAs heterostructures.\cite{han01} Thus, a decay $\propto \exp{(-cT)}$ of the AB amplitude is expected below $T_c$ which is in reasonable agreement with our 
observation in Fig.~2(e).  Due to a limited accessible temperature range we can not extract $T_c$ from the data and determine the energy scale, when thermal averaging contributes to decoherence of AB oscillations.

For further analysing the AB oscillations we preform fast Fourier
transformation (FFT) of the processed data, as shown exemplarily for
$V_G=3$\,V in Fig.~3(a). The Fourier spectrum shows a broad band of
frequencies that constitute the signal. Peaks for frequencies below the
expected AB oscillations, which we attribute to artifacts from the data
averaging of a $15$~mT moving window (corresponding to $66$~1/T) are neglected
in the further discussion.  The interval of the $h/e$ mode frequencies is
given by the sample geometry ranging from $125$ to $433$~1/T as indicated in
Fig.~3(a). The main contribution is around the frequency related with the mean radius $\overline{r}$ of the ring, although features at lower and higher frequencies are present. The broad spread is due to the aspect ratio of mean radius $\overline{r}$ and the width $w$ of the ring, $\overline{r}/w \sim 1.6$. 
Notably, this circumstance makes any attempt to distinguish contributions from the $h/e$
and $h/2e$ mode difficult as the differences in enclosed area for different
paths overlap (see also horizontal arrows in Fig.~3(a)). 
This experimental observation is also well reproduced by 
 tight binding calculations as describe above (see Fig.~3(b)). In the simulations, the AB oscillations exhibit a higher amplitude than in the experiment, which we attribute to the perfectly symmetric transmission through both ring arms and the lack of any other experimental limitations, such as bulk disorder and contact resistance mentioned above. These effects are not included in the calculations since they are not relevant to the investigated physics. 
The overall behavior of the conduction fluctuations in the wider ring is
comparable to the experimental data and the FFT analysis gives similar results
(compare black traces in Figs.~3(a) and 3(c)). Again, a broad range of frequency
components is observed in the given ranges of AB mode making a clear
assignment of FFT peaks to mode number more than difficult. The narrower ring
with a larger aspect ratio ($\overline{r}=700$~nm, $w=200$~nm) exhibits a
lower amplitude $\Delta G_{AB}$ in the calculations, but more distinguishable
peaks in the FFT spectrum (compare traces in Fig.~3(c)). Now peaks only appear
close to the center of the corresponding mode ranges and are unambiguously
identified as the fundamental mode and its multiples. Since decoherence, i.e.
dephasing is not implemented in the simulation, many higher harmonics are
visible. To confirm these results we measured AB oscillations in sample \#2
which has a larger aspect ratio $\overline{r}/w \sim 2.5$. Figure~3(d) displays
the FFT spectrum of sample \#2 at similar $n$ as Fig.~3(a). The ranges of $h/e$ and $h/2e$ conduction oscillations only overlap slightly and peaks of the fundamental mode and the first and second harmonic are observed. In particular, the fact of observing contributions $h/3e$ conductance oscillations (see label in Fig.~3(d)) allows for the conclusion that the phase-coherence length $l_{\phi}$ exceeds the value of $\approx 1.5\, \mu m$. 
Hence the difficulties of AB mode identification for sample \#1 arise from the ring geometry and does not primarily reflect on the sample quality and the phase-coherence length.

\subsection{Ballistic electron guiding and magnetic focusing}
\begin{figure*}[tb]\centering 
\includegraphics[draft=false,keepaspectratio=true,clip,%
                   width=0.95\linewidth]%
                   {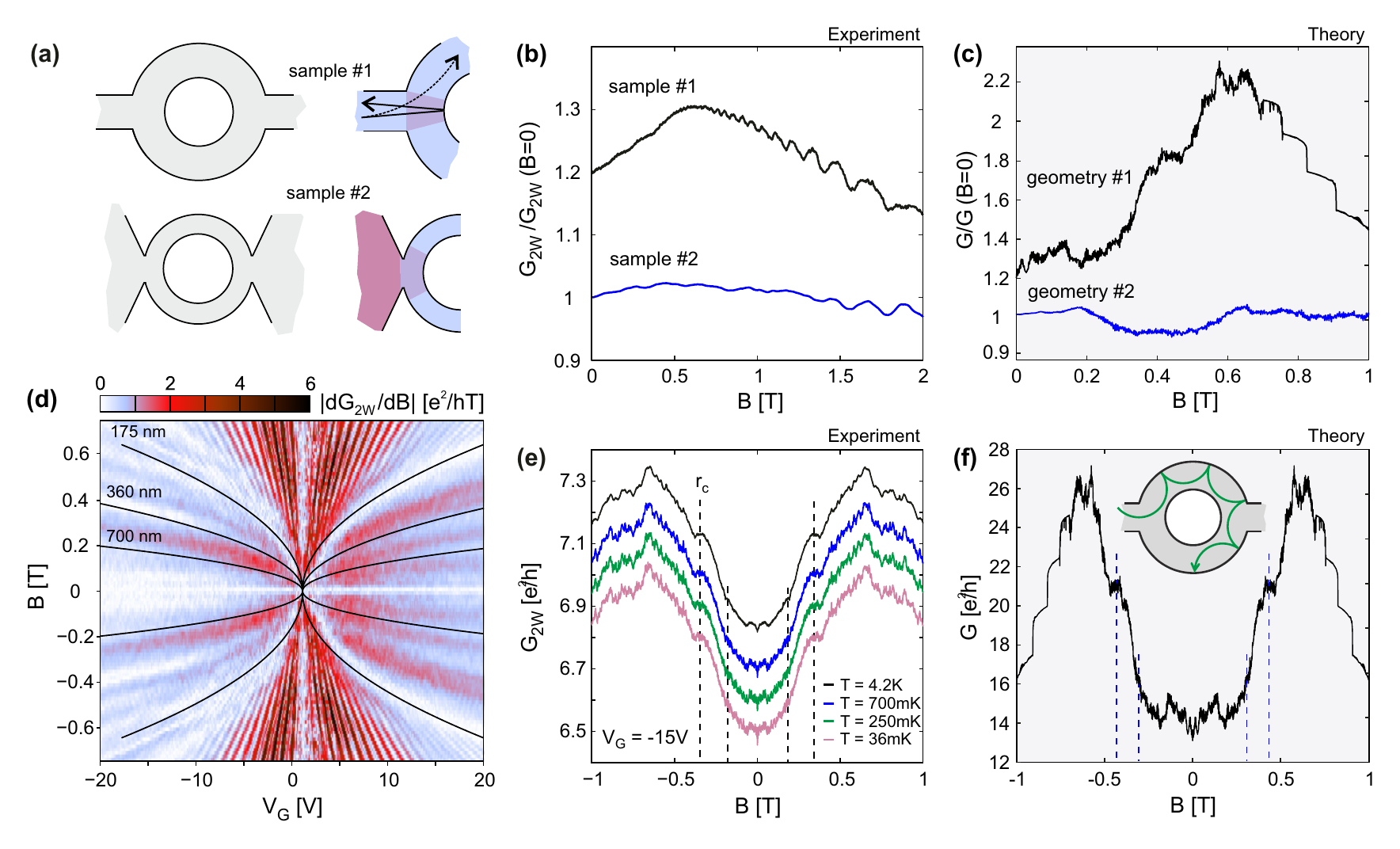}                   
\caption{
(a) Schematic representation of the different ring geometries of the samples
\#1 and \#2. (b) Magnetoconductance for both samples \#1 and \#2 in the
intermediate magnetic field regime for similar $n \approx 1.3\times
10^{12}$\,cm$^{-2}$ (for sample \#1, $V_G=20$\,V). Values are normalized with
respect to the conductance at zero $B$-field and an offset is added for
clarity. (c) Simulated magnetoconductance $G$ for different device geometries
\#1 and \#2 similar to (b) for $n=1\times10^{12}$\,cm$^{-2}$ and $\delta
r=16$\,nm (geometry \#1: $\overline{r}=580$\,nm, $w=350$\,nm; geometry \#2:
$\overline{r}=500$\,nm, $w=200$\,nm; plotted with an offset). (d) Derivative of
two-terminal conductance with respect to magnetic field $dG_{2W}/dB$ as
function of $V_G$ and $B$. Solid lines highlight constant cyclotron radii for
selected values. (e) Two-terminal conductance $G_{2W}$ versus magnetic field
$B$ for $T=36$\,mK and $4.2$\,K at $V_G=-15$\,V
($n=1\times10^{12}$\,cm$^{-2}$). Curves are plotted with an offsets for clarity.
Vertical dashed lines represent a cyclotron radii of $r_{C}=360\,$ and
$r_{C}=700$\,nm. (f) Conductance $G$ versus magnetic field $B$ of a ring with
geometry \#1 calculated by tight binding approach also for
$n=8.5\times10^{11}$\,cm$^{-2}$ and $\delta r=16$\,nm. Vertical dashed lines
again represent cyclotron radii as depicted in panel (d). Inset illustrates the trajectory of charge carriers inside a conductance plateau.
} 
\label{figure4}
\end{figure*}
For magnetic fields of up to $B=0.5$~T $(w\approx r_C)$ we observe an increase of the magnetoconductance before a decrease evolves while entering the quantum Hall regime. In Figs. 4(a) and 4(b) this effect is shown for the two samples at similar charge carrier density $n=1.3\times10^{12}$\,cm$^{-2}$ but different geometries, which differ in the type of connecting the ring structure to the source and drain leads and the width of the ring arms. Sample \#1 consists of a kind of
a T-junction comparable to geometries used in previous experiments for
graphene rings on silicon dioxide~\cite{rus08,hue09,hue10} while sample \#2 is
based on a V-shaped connection similar to experiments on III-V
heterostructures~\cite{grb07}. The changes in $G_{2W}$ with $B$ are more
pronounced in sample \#1 with the T-junction, while they are hardly visible in
sample \#2 (see Fig.~4(b)).
The increase in magnetoconductance for sample \#1 is caused by an increase of the average mode transmission 
as function of $B$-field at the T-junction. From a semiclassical point of view this observation is connected to the fact that straight trajectories ($B = 0$~T) are more likely reflected at the T-junction than curved once ($B \neq 0$~T), as illustrated in Fig.~4(a) (see solid and dashed trajectories in the upper right panel). This effect, however, is strongly reduced in 
sample \#2 because there are many open modes available in the lead region very close to the ring, making transport not very
sensitive to the average mode transmission as the overall conductance is
limited only by the number of modes in the ring arms (compare different colors
in the right panels of Fig.~4(a)). In other words, there are in any case enough trajectory angles available for entering the ring via the V-shaped connection. 
These observations support the assumption of quasi-ballistic transport
($w<l_m<l$), because magnetic focusing requires a mean free path $l_m$ larger
than the width $w$ of the ring. In simulations with similar device geometries
--- geometry \#1 (T-geometry with $\overline{r}=580$\,nm, $w=380~nm$) and
geometry \#2 (V-geometry with $\overline{r}=500$\,nm, $w=200$\,nm) and
comparable charge carrier density $n=8.5\times10^{11}$\,cm$^{-2}$ --- we find
good qualitative agreement with our experimental findings (see Fig.~4(c)). The simulations show an increase in magnetoconductance before entering the quantum Hall regime for the T-junction geometry (\#1) and an almost constant curve for the V-shape geometry (\#2). 
 Steps indicating the beginning of the quantum Hall effect occur for geometry
 \#1 at $B \approx 0.7$\,T, whereby for the narrower geometry \#2 they start
 at $B=1.2$\,T since a smaller cyclotron radius is required (not shown).
 Intrinsic restrictions of the simulations, such as zero temperature, infinite
 phase-coherence length and negligence of contact resistance, lead to
 additional features. At zero temperature UCFs are enhanced leading to more wobbly traces. Also Shubnikov-de Haas oscillations are not visible due to the absence of thermal broadening of the Landau levels. In general the simulation results coincide with the experimental observation and the effect is attributed to the interplay of focusing of charge carriers by magnetic field and the type of connection between ring and leads, which is refereed to as a size effect. In these measurements we find also plateau-like features in the rising magnetoconductance below $B\leq 0.5$\,T indicating another size effect related with magnetic field. 

Figure 4(d) depicts the derivative of conductance with respect to magnetic
field $dG/dB$ as function of $V_G$ and $B$ in a color plot. Here, several
features are identified, which show a dependence as the cyclotron radius
$r_C(B,V_G) = \hbar \sqrt{\pi \alpha V_G}/(e B)$. As a guide to the eye
constant cyclotron radii are drawn in Fig.~4(d) using the gate lever arm
$\alpha$ determined from Landau fan measurements (see below). At $r_C =
175$\,nm the cyclotron radius matches the requirement for the formation of
edge states ($r_C \leq w/2$) and we observe the transition to the quantum Hall
effect exactly at this limit. In the regime of larger cyclotron radii ($r_C >
w/2$) we interestingly observe constant conductance plateaus ($dG_{2W}/dB=0$)
in the $B$-field dependent increasing magnetoconductance, which coincide with
$r_C \approx 360$ and $700$\,nm, respectively. For further investigations
$G_{4W}$ versus $B$ is measured at fixed gate voltage $V_G=-15$\,V ($n\approx
1\times10^{12}$\,cm$^{-2}$) for different temperatures $T=36$\,mK, $250$~mK,
$700$~mK and $4.2$\,K (see Fig.~4(e)). For all temperatures the traces are very similar and only differ in the amplitude of the UCFs and AB oscillations, which almost vanish for $4.2$~K. At $B=350$\,mT a conductance plateau is observed, which corresponds to a cyclotron radius of $r_{C}=360$\,nm. A second conductance plateau is located at $B=182$\,mT (corresponding to $r_{C}=700$\,nm), but only visible at $4.2$\,K due to the superposed AB oscillations at lower temperatures. Since the plateaus are observable even in the absence of UCFs and AB oscillations, we conclude that they do not require phase-coherent transport. 

The high carrier mobility and the dependency on cyclotron radius indicate that the conductance plateaus are linked to magnetic focusing of charge carriers inside the ring structure.\cite{tay13} In this case the trajectories of the carriers are deflected by the magnetic field and for some cyclotron radii it becomes more unlikely to exit the ring, which yield in a suppressed increase of magnetoconductance. 
A schematic illustration of an example of such a trajectory is shown in the
inset of Fig.~4(f). For confirming our observation, we carried out tight
binding calculation in the intermediate magnetic field regime. Figure~4(f) shows
corresponding simulations for the data shown in Fig.~4(e). Here the conductance plateaus are reproduced quantitatively (compare vertical dashed lines in Figs.~4(e) and 4(f)) by assuming finite edge roughness $\delta r$ of the graphene ring. The edge roughness introduced by the lithographic pattering of the graphene sheet affects the visibility of the focusing effect and a further analysis of its impact on the simulation results can be found in Appendix~3. 
The smooth steps in the falling edge after the maximum conductance are due to the emerging quantum Hall effect and are not related to magnetic focusing or size quantization effects. Similar observations of magnetic focusing have been made in an AB ring in a GaAs quantum well, where also a T-junction has been used and these effects have been attributed
 to quasi-ballistic transport with boundary scattering.\cite{mur08} 
This type of boundary scattering is quite surprising in consideration of the lithographically etched sample outline and the requirement for specular scattering, that the roughness of the boundaries must be smaller than the Fermi wavelength $\lambda_F$.\cite{bee91} Thus the spatial extension of the disorder potential from the device edge must be smaller than $30\,\text{nm}$, since the focusing feature is still present at gate voltage $|V_G|=20\,\text{V}$, which correspond to $n=1.4 \times 10^{12}\,1/\text{cm}^2$ and $\lambda_F \approx 30\,\text{nm}$, respectively.
In summary, our experimental findings and simulation results point in exactly this direction and strongly support the observation of magnetic focusing in a quasi-ballistic 
ring system.
\subsection{High-visibility AB oscillations near QH plateaus}
\begin{figure*}[tb]\centering 
\includegraphics[draft=false,keepaspectratio=true,clip,%
                   width=0.95\linewidth]%
                   {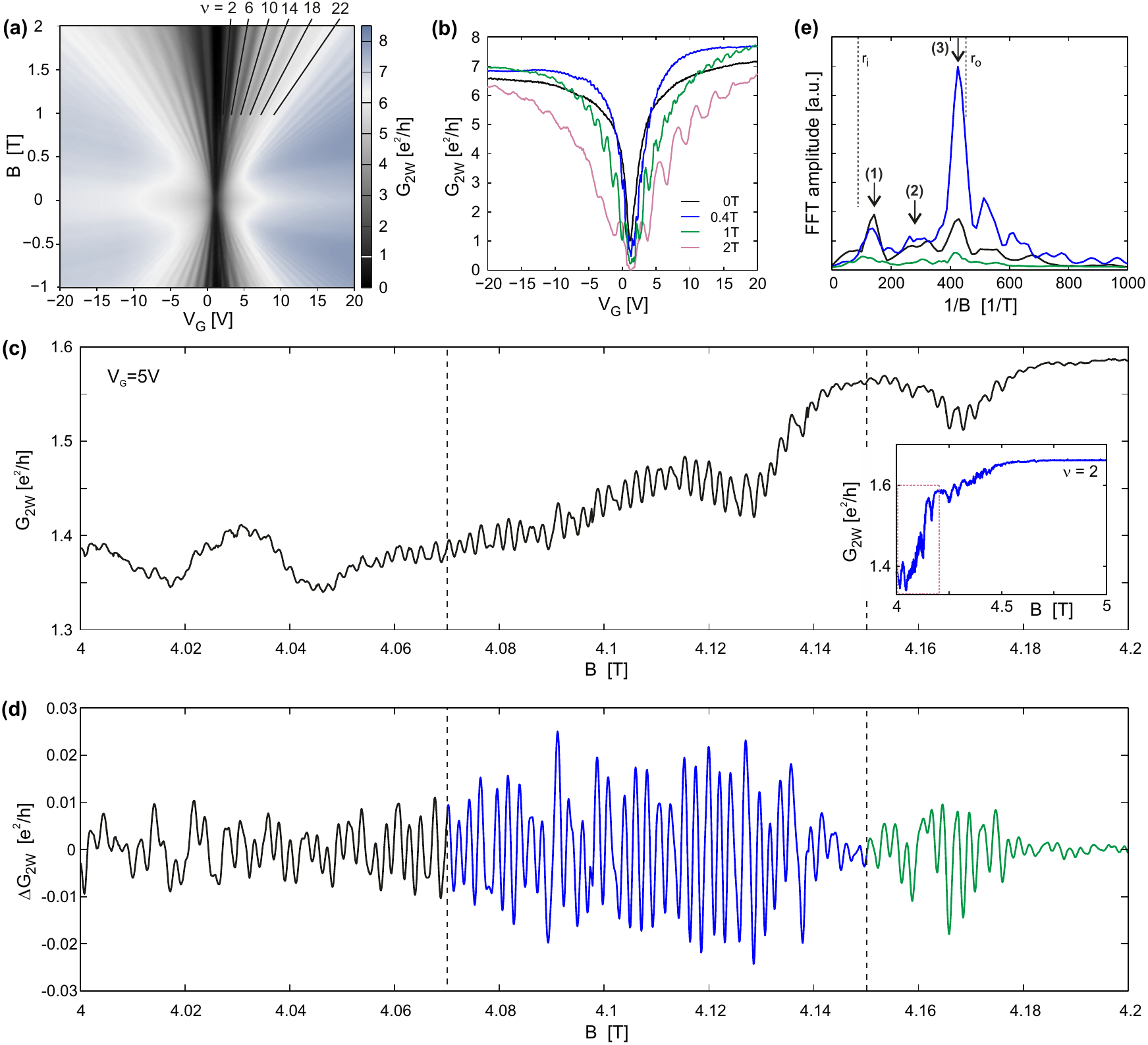}                   
\caption{
(a) $G_{2W}$ as function of $V_G$ and $B$ at $T=36$\,mK. Solid lines represent slopes used for the extraction of $\alpha$.  (b) Line cuts of $G_{2W}$ versus $V_G$ for various $B$ extracted from panel (a). (c) Zoom-in of high resolution measurements of $G_{2W}$ as function of $B$ at $V_G=5$\,V in the slope before quantum Hall plateau $\nu=2$. Inset shows larger measurement range. Red box indicate the selected $B$-field region. (d) Background subtracted $\Delta G_{2W}$ as function of $B$ from data shown in panel (c). (e) Fourier spectra of $\Delta G_{2W}$ from different regions as depicted in panel (d).
} 
\label{figure5}
\end{figure*}

\begin{figure*}[tb]\centering 
\includegraphics[draft=false,keepaspectratio=true,clip,%
                   width=0.95\linewidth]%
                   {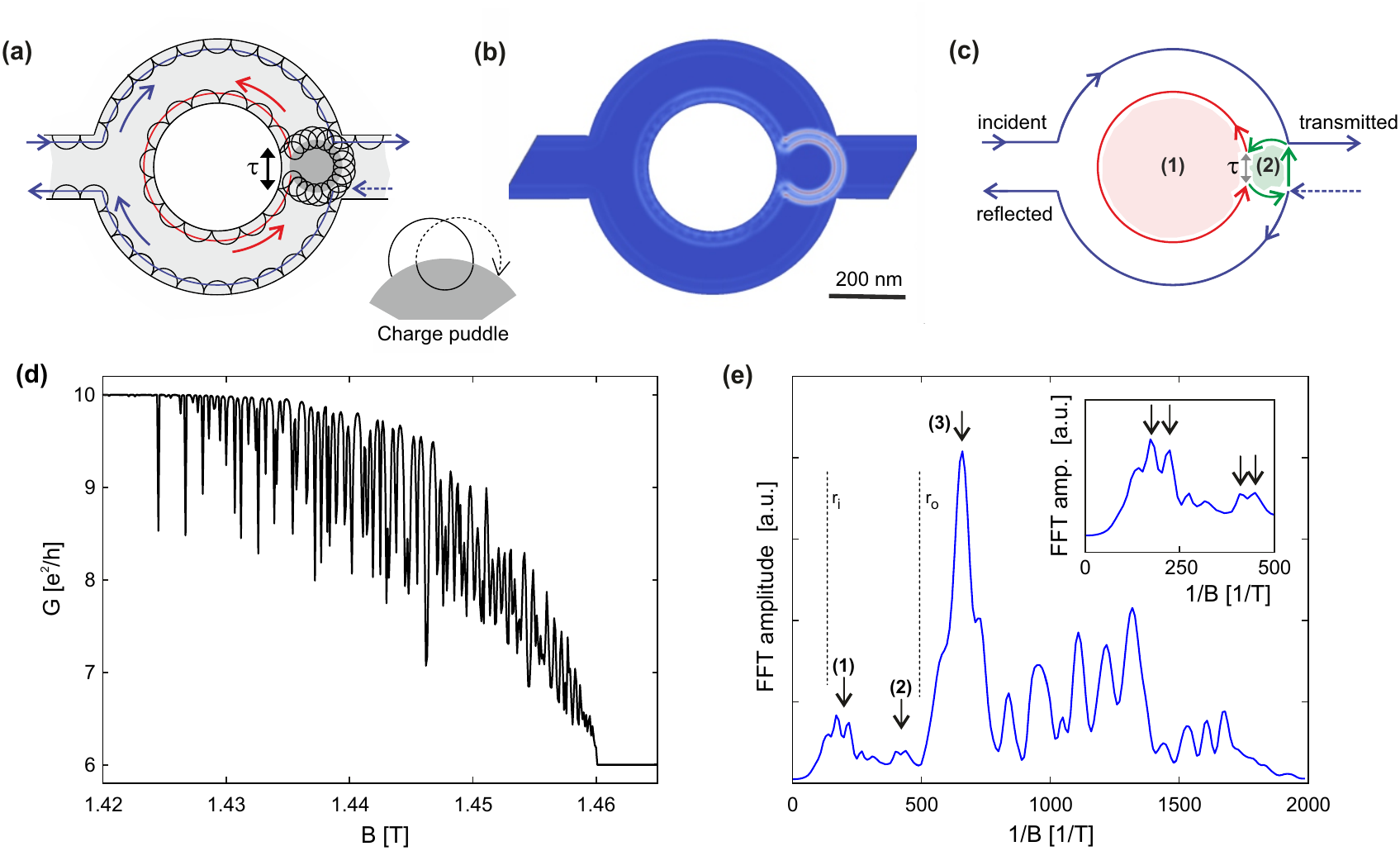}                   
\caption{%
(a) Schematic illustration of a ring with a charge puddle connecting the inner and outer edge
channel in the quantum Hall regime. Arrows indicate the direction of the edge channels. The inset highlights cycloid drift motion of an edge channel along the charge puddle. (b) Calculated density of states of a ring
in quantum Hall regime with $n=2.8\times10^{11}$\,cm$^{-2}$ and $\delta r=0$
including a charge puddle with $d_P=350$~nm and $\Delta
n=-1.6\times10^{10}$\,cm$^{-2}$. (c) Schematic illustration of the edge states
	contributing to the interference: the path of the electron that is directly
	transmitting interferes with the path that encircles the green and red
	disk (following the arrows). This interference can be tuned via the AB phase
	of the area (green and red) encircled. Note that in order for interference
	to happen at all, part of the wave function has to leak to the reflecting edge channel as otherwise unitarity ensures perfect transmission. (d) Calculated conductance $G$ as function of magnetic field between two quantum Hall plateaus based on
density of states shown in panel (b). (e) Fourier spectrum of background subtracted calculated conductance.
The inset shows a close-up of the FFT spectrum. For more information see text.
} 
\label{figure6}
\end{figure*}

Next we focus on the high $B$-field regime ($r_C<w/2$).
Figure 5(a) shows a 2D color plot of $G_{2W}$ as function of $V_G$ and $B$. We
observe a graphene-typical Landau fan for two-terminal measurements with
integer Hall plateaus and filling factors up to $\nu=\pm22$ for $B \leq 2$\,T.
The observation of clear and pronounced features in the Landau fan at such low
magnetic fields presupposes a very homogenous doping level, which is also
reflected in the low extracted $n^\ast$ (see above). Following Ref.~\onlinecite{ki12} we extract the capacitive coupling or so-called
lever arm $\alpha =\Delta n / \Delta V_G$ of the back gate from the slope of the Landau levels, which results in $\alpha=6.7\times10^{10}$\,cm$^{-2}$V$^{-1}$ and is in good agreement assuming a parallel plate capacitor model. 
 
In Fig.~5(b) we show several traces of $G_{2W}$ as function of $V_G$ for various $B$-fields. Quantum Hall signatures are evolving in the vicinity of the CNP below $B=0.3$\,T and are fully developed at $B=0.5$\,T. The dips in magnetoconductance between the conductance plateaus are due to the sample geometry $w<l$ as described above. For $B=2$\,T the conductance plateaus are more pronounced, but also an asymmetry between the hole and electron transport regime becomes apparent, whereas the origin of this remains unclear.

In the $B$-field range of fully developed quantum Hall plateaus, we find a
substantial increase in the visibility of AB oscillations for specific
settings of $B$ and $V_G$ (see Fig.~5(c)). This effect takes place at the
flanks, i.e. onsets of quantum Hall plateaus and is observed only for a few
different $B$-field and gate voltage values. The AB amplitude $\Delta G_{AB}$
changes by a factor of $2-3$ for small changes in $B$-field on the order of a
few tens of milliteslas indicating that they are sensitive to sample
inhomoginities. Figure 5(d) depicts $\Delta G_{2W}$ of the data shown
Fig.~5(c). In the best region the AB amplitude reaches a visibility of more
than $0.7\%$, which is a significant increase compared to the low field regime
($\approx 0.25\%$). This becomes even more apparent in the FFT spectra of the
different ranges. Figure 5(e) shows the corresponding FFT spectra and we observe an increase of the FFT amplitude by a factor of three or more.
The distinct peaks in the FFT spectra (see e.g. blue trace in Fig.~5(e)) moreover show that there are very distinct enclosed areas contributing to the AB oscillations. This is overall in line with edge channel dominated transport. Interestingly we find (i) that there is one conductance oscillation frequency (labeled with (1) in Fig.~5(e)) which can be associated with the area enclosed by an edge mode along the inner radius $r_i$ (see vertical dashed line in Fig.~5(e) and red circle in the schematic illustration in Fig.~6(a)) and
(ii) that the most pronounced peak in the FFT 
can be hardly associated and explained with an edge channel traveling along the outer radius $r_o$ (see vertical dashed line). 
This is on one hand because peak (3) includes area contributions larger than
$\pi r_o^2$ and on the other hand it is impossible to close such an area by
pure edge channel transport. 
The latter point is also crucial when considering mechanisms which allow to access (i.e. to connect to) an edge channel propagating along the inner edge of the ring structure (see red circle in Fig.~6(a)).

Consistent with the observation that the conductance oscillations are only present in the crossover regime from one to another filling factor we assume that the disorder potential in the ring entrance or exit region is playing an important role. A present disorder landscape indeed may allow for both (i) accessing the inner edge channel and (ii) connecting the upper and the lower ring arm (see Fig.~6(a)). Such a disorder potential is known to arise from
substrate interaction, contaminations or rough edges.\cite{mar07}
It is very likely that in our case there is just one dominating "`charge puddle"' connecting the two edge states, as scanning tunneling microscopy studies have shown that in graphene on hBN such disorder induced puddles have typical diameters of around $100-200$\,nm,\cite{xue11} which is on the order of $w$.

For a more detailed discussion we performed calculations of magnetotransport through an
ideal ring structure (no edge roughness, $\delta r=0$) with similar dimensions
($w = 400$~nm, $\overline{r} = 600$~nm) and include one charge puddle with a
diameter of $d_p = 350$~nm located at the exit region of the ring (see Fig.~6(a)).
In Fig.~6(b) the calculated local density of states in the quantum Hall regime
is plotted for a charge carrier density of $n=2.8\times10^{11}$\,cm$^{-2}$
including a charge puddle with an offset in charge carrier density of $\Delta
n=-2\times10^{10}$\,cm$^{-2}$ sitting right in the exit region of the ring
(see Figs.~6(a) and 6(b)). Using this charge carrier density for Kwant
simulations of magnetotransport the effect of enhanced visibility of AB
oscillations is reproduced in qualitative manner (see Fig.~6(d)), where AB oscillations become most visible between two conductance plateaus (in this example at the step from filling factor 10 to 6). 
Again, many oscillations are observed with a maximum amplitude in the middle
between two quantum Hall plateaus. In Fig.~6(e) we show the corresponding Fourier spectrum of the background subtracted conductance and the spectrum exhibits a surprisingly similar peak structure as seen in the experiment (see labels (1), (2) and (3) in Figs. 5(e) and 6(e)). 
In Fig~6(e) we can again identify a frequency contribution corresponding to
the area enclosed by the quantum Hall channel around the inner radius of the
ring (see left vertical dashed line and label (1) in Fig.~6(e)) while the most pronounced peak is at higher frequency value, exceeding an enclosed area of $\pi r_o^2$ (see right vertical dashed line).
By making use of our model system we can explain the enclosed area leading to the main peak
(3) as being the sum of the two areas (1) and (2).
The first area (1) corresponds to the edge channel going around the inner ring. The second area (2) originates from the motion around the charge puddle (see schematic illustrations in Figs.~6(a) and 6(c)).
Notably, the 
second contribution (see also labels (2) in Figs. 6(e) and 5(e)) gives rise to
a higher frequency compared to the one due to the inner ring radius even though 
its geometrical area is smaller. This is due to the different confinement of the edge
channel (see schematic illustrations in Fig.~6(a)). 
While the confinement due to the graphene edges is a hard confinement with the edge modes corresponding to classical skipping orbits, the confinement along the charge puddle is a soft confinement corresponding to  a cycloid motion (full cyclotron orbits that are drifting as shown by the inset in Fig.~6(a)).
In a semiclassical description, the cycloid motion encloses a much larger area
than the area enclosed by the guiding center.  In particular, in a simplified
model of the electron drift, the charge puddle has a different but constant
filling factor and half of each cyclotron orbit is on either side of the
interface. A single cyclotron orbit encloses an area of $A_C= \pi r_C^2$. The
drift $\Delta l$ is given by twice the difference of the cyclotron radius,
i.e., $\Delta l = 2 \Delta r_C \approx r_C \Delta n /n$ where $n$ is the
charge carrier the density and $\Delta n$ the difference in carrier densities
of the bulk and the charge puddle. Thus, the electron motion encloses the
effective area  $A_{\text{eff}} = A_C L/\Delta l= \pi L r_C n /\Delta n$ due
to the cyclotron motion, where $L$ is the length of the quantum Hall edge
channel, additional  to the geometric area of the charge puddle enclosed by
the guiding center. In the present case, $A_{\text{eff}}$ even dominates over
the geometric size of the charge puddle resulting in the surprising effect
that the peak (2) is due to the (small) charge puddle whereas (1) is due to
the (large) inner radius of the ring. We have numerically checked that the
dominant Aharonov Bohm frequency of the charge puddle alone without inner ring
is $400$~1/T that corresponds to the peak (2). Including the effect of the
inner ring, the dominant contribution (3) is due to electrons encircling both
the charge puddle as well as the inner ring. However, due to a finite
tunneling coupling $\tau$ (see illustrations in Figs.~6(a) and 6(c)) along the
charge puddle and the boundary of the inner ring, also the fundamental
frequencies (1) and (2) contribute slightly to the Fourier transform.
Interestingly, by having a closer look at the corresponding peaks in the FFT spectrum (see inset in
Fig.~6(e)) we observe a double peak structure (see arrows therein). This peak
``splitting'' is due to
the valley degeneracy lifting of the edge modes and is also well observed in the calculated density of 
state map shown in Fig.~6(b). See in particular at the white double ring
structure around the charge puddle (compare Figs.~6(a)--(c)).

The visibility of all these effects in the simulation depends heavily on the specific magnetic field range, charge carrier density, puddle size, and offset in Fermi energy.  However, they can be adjusted for all cases in order to observe the enhanced visibility. Since some of these parameters are unalterable in the measured sample, the rare observations of this effect becomes fairly reasonable.

\section{Conclusion and Summary}
In conclusion, we report on the fabrication and characterization of graphene rings encapsulated in hexagonal boron nitride. We demonstrate high carrier mobilities and show magnetotransport measurements exhibiting fully phase-coherent transport. In particular, we observe the fundamental mode ($h/e$) and higher modes ($h/2e$ and $h/3e$) of Aharonov-Bohm conduction oscillations. In the intermediate $B$-field regime, we identify signatures in the magnetoconductance resulting from electron guiding and magnetic focusing of charge carriers indicating quasi-ballistic transport. These experimental observations are reproduced by tight binding simulations verifying quasi-ballistic transport and magnetic focusing. 
Finally we discuss the observation of AB oscillation in the quantum Hall regime,
where at the cross-over from adjacent filling factors AB oscillations with a visibility on the order
of 0.7\% are measured. By a detailed analysis we can attribute these oscillations to the sum of the 
areas enclosed by edge channels along the inner ring radius and a charge puddle connecting the inner and
outer ring edge. Interestingly, the corresponding interference path, i.e. the edge mode propagates along segments of hard as well as soft-wall confinement.

Overall our work shows that graphene-hBN sandwiches serve as an interesting
host material for studying mesoscopic physics. In particular in view of the
high tunability of the Fermi wave length graphene may  open very interesting avenues for investigating more complex interferometers, quantum billiards, Andreev billiards with hard wall confinement as well as Dirac fermion optic devices.  



\section{Acknowledgments}
We thank S. Engels and B. Terr\'{e}s for help on the fabrication process and
H. Bluhm and M. Morgenstern for fruitful discussions. Support by the Helmholtz
Nanoelectronic Facility (HNF)\cite{HNF17} at the Forschungszentrum J\"ulich,
the Excellence Initiative of the Deutsche Forschungsgemeinschaft (DFG), the EU Graphene Flagship project
(contract No. 696656), and the ERC (contract  no. 280140) are gratefully acknowledged. Growth of hexagonal boron
nitride crystals was supported by the Elemental Strategy Initiative conducted by the MEXT, Japan and JSPS KAKENHI Grant Numbers JP26248061, JP15K21722 and JP25106006.

\appendix*

\section{Comparison to numerics}

\begin{figure*}[tb]\centering 
\includegraphics[draft=false,keepaspectratio=true,clip,%
                   width=0.95\linewidth]%
                   {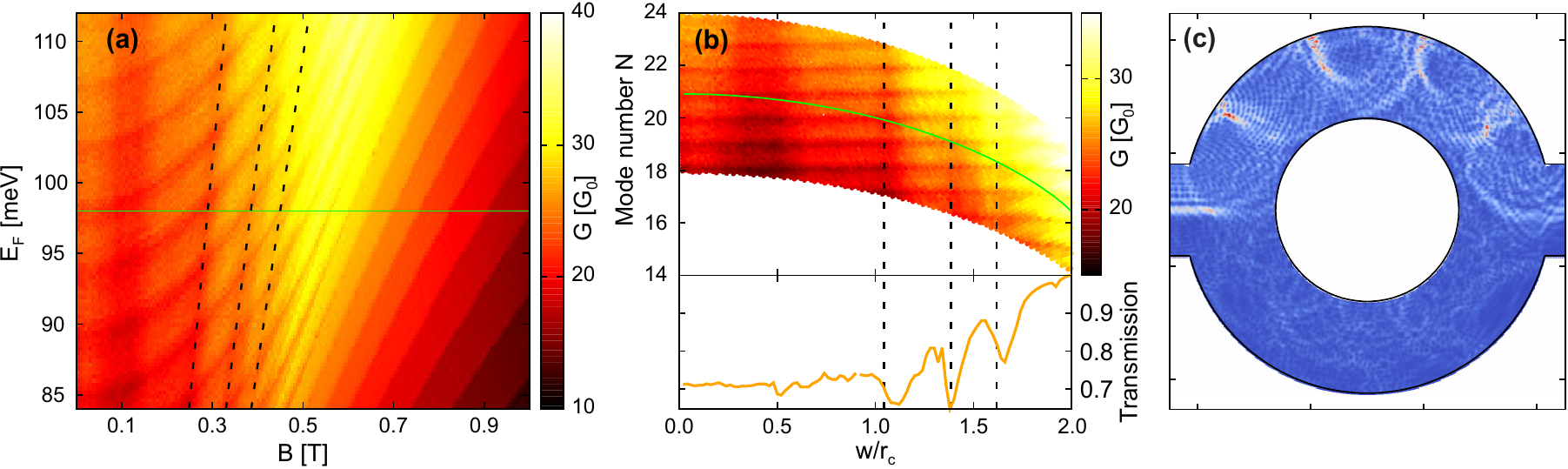} 
	\caption{(Color online)%
(a) Conductance in units of $G_0 = 2e^2/h$ as a function of the magnetic field
$B$ and the chemical potential (Fermi energy)  $E_F$ obtained by a tight-binding calculation of a graphene ring with inner
radius $r_{i} = 400$\,nm and outer radius  $r_{o} = 800$\,nm. Indicated with
dashed lines are lines of fixed cyclotron radius $r_C = 429\,\text{nm},
569\,\text{nm}, 665\,\text{nm}$.
(b) Conductance data of (a) replotted as a function of the number of
modes $N$ propagating in the ring (of width $w= 400$\,nm, see text) and the inverse cyclotron
radius $1/r_C = v_F e B / E_F$ in units of the inverse width of the ring $1/w$. Minima and maxima
of the conductance are approximately horizontal and vertical on this plot. The solid green
line is the line of constant energy along which Fig.~\ref{fig:numericsb}(c) is evaluated.
The lower panel shows the semiclassically calculated transmission through the ring (for more details see text).
(c) Plot of the density of states of an almost perfectly transmitting mode.
We observe that the trajectory of the electron starting in the
left lead performs a skipping orbit which after four reflections at the
boundary enters the right lead.	}
\label{fig:numericsa} 	
\end{figure*}

\begin{figure}
\includegraphics[draft=false,keepaspectratio=true,clip,%
                   width=0.62\linewidth]%
                   {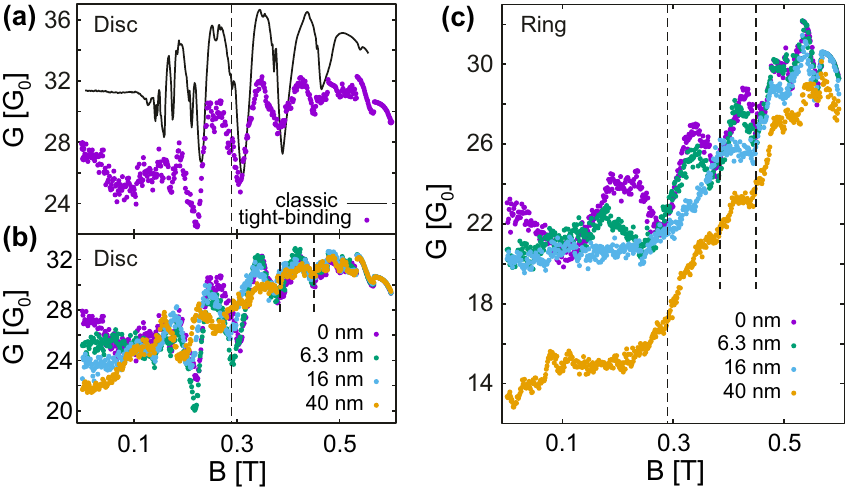} 
\caption{(Color online)%
(a) A comparison of the tight-binding result for the
transmission to a simple classical description at a Fermi Energy $E_F = 98\,$meV for the transport through a disk ($r_{i} = 0$). 
A correspondence of the features in the two calculations is visible all the way
up to the quantum Hall limit $w = 2 r_C$. Note that the results are also
approximately match the results of Fig.~\ref{fig:numericsa}(b) for $w\gtrsim r_C$ indicating
that the inner radius has only an effect for small cyclotron radii.
(b) Simulation of the disc geometry with a sinusoidal modulation of
the outer radius with amplitude $\delta r= 0,\, 6.3,\,16,\, \text{and}\, 40\,$nm. The conductance for
the disk is shown for different strength of edge roughness with the result
that the position of the conductance minima are rather robust to edge roughness.
(c) Finally, we also include edge roughness in the simulation of the graphene ring.
We observe that with increasing edge roughness the features of
quantization and magnetic focusing weaken until they resemble a shoulder like
structure that was observed in the experiments.
\label{fig:numericsb}
}
\end{figure}

\subsection{Implementation}
The quantum transport simulation have been performed using a tight-binding
approximation on a hexagonal lattice using the Kwant package.\cite{kwant}  We
have simulated a scaled version of the graphene lattice with a lattice
constant that is a factor 10 larger than the experimental situation. For ease
of comparison, all the results are presented in the original units; in
particular, the graphene sheet is described by its Fermi velocity  
  $v_F = 1.15
\times 10^6\,\text{m}/\text{s}$ and the electron
density $n$  implemented via the Fermi energy $E_F = \hbar v_F \sqrt{\pi n}$
measured with respect to the Dirac point.  We have simulated a ring with inner
radius of size $r_{i} =400\,$nm and outer radius $r_{o}=800\,$nm
contacted by two leads of width $w=400\,$nm as illustrated in
Fig.~\ref{fig:numericsa}(a). Following Ref.~\onlinecite{wurm}, we have
implemented the leads by infinitely extended waveguides of highly doped
graphene.  The leads are connected to the device by an impedance matching zone
of length 200\,nm where the Fermi energy is continuously reduced from the
large value in the leads to the value $E_F$ in the device. In total, our
simulation involves $7 \times 10^5$ lattice sites.

\subsection{Conductance steps and magnetic focusing}
We present the conductance of the system as a function of the homogeneous magnetic field $B$ and the Fermi energy $E_F$ in Fig.~\ref{fig:numericsa}(a). We observe multiple lines of local minima in this plot. In order to investigate their origin, we have replotted the data again in Fig.~\ref{fig:numericsa}(b) with different axis. On one axis, we have placed a semiclassical approximation for the number of modes in a graphene ribbon of the width $w$ of the lead that connects to the ring (also equal to the width of the ring $w =r_{o} - r_{i}$). The number of modes $N$ is evaluated as a function of the two geometric scales in the setup, the magnetic length $l_B^2 = \hbar / e B$ and the Fermi wavenumber $k_F = E_F / (\hbar v_F)$ and is given by
\begin{equation}
N = \frac{k_F^2 l_B^2}{\pi} \text{arcsin}\Big(\frac{w}{2 k_F l_B^2}\Big) +
\frac{w}{4 \pi} \left(4 k_F^2 - \frac{w^2}{l_B^4}\right)^{1/2} \label{nsemi}.
\end{equation}
The expression is obtained using the translational symmetry of the problem to reduce it to one dimension and then using WKB for an approximate solution. \cite{semiribbon} On the other axis we have used the inverse cyclotron radius $r_C^{-1} =  v_F eB/E_F$ in units of the inverse width $w^{-1}$ of the geometry.

With the new axes, there are horizontal lines as well as vertical dips visible. We attribute the horizontal features to conductance quantization as they occur where the appearance of a new mode is expected in a plain graphene ribbon.  The vertical features are more intriguing and are the main focus of this section. They occur at a fixed ratios of the cyclotron orbit $r_C$ with respect to the geometry of the ring. Therefore, it seems likely that they are connected to magnetic focusing due to the bending of the classical trajectories of the electrons in a magnetic field.\cite{houten}

In order to test this hypothesis, we have performed a set of classical simulations of the transport through the ring. We have traced the path of individual trajectories propagating through the geometry assuming specular reflection at the boundary. With each trajectory starting from a random initial condition in the lead that is propagating toward the ring. We evaluated in a Monte-Carlo scheme the fraction $T$ of trajectories that propagate through the ring. In order to compare the findings with the quantum results we evaluated $G_{\text{classic}} = T N G_0 $, with $N$ being the semiclassical number of modes as in Eq.~(\ref{nsemi}).

Comparing the position of the vertical dips with the dips in the transmission probability $T$ in Fig.~\ref{fig:numericsa}(b), we observe a correspondence for $w \gtrsim r_C$. Note that when keeping the Fermi energy constant, we follow a bent curve (e.g., the solid green line for $E_F = 98\,$meV) and thus in traces we can see both the horizontal as well as the vertical features.

Fig.~\ref{fig:numericsa}(c) displays the electron density of a specific mode that is incoming
from the left lead. The maximum of the density follows semicircular paths that
correspond to classical cyclotron orbits. The ratio of the cyclotron radius and the dimension of the geometry is chosen such
that after four (almost specular) reflections at the upper part of the
outer radius, the electron enters the right lead. Such a mode contributes to
transport with a conductance close to the maximal value $G_0$. Increasing
the magnetic field slightly decreases the cyclotron radius
such that at some point there will be a reflection right before entering the
right lead and the trajectory of the electron subsequently skipping over the
right lead. Here, the probability for entering the right lead is
heavily lowered which results in a reduction of the conductance. As a
result the conductance oscillates as a function of the magnetic field. This is
the reason for the magnetic focusing shown in Fig.~\ref{fig:numericsa}(c).

In Fig.~\ref{fig:numericsb}(a), we have performed  a more detailed comparison
for a simpler setup  of a graphene disk with $r_{i}=0$. The idea is that the
inner radius is not important at all for the occurrence of the effect. In
particular, no well-defined trajectories imposed by the constriction  are
required as in the case for Aharonov-Bohm oscillations. The results of the
simulation are in good agreement. In particular, the classical prediction
for the transmission probability $T$ matches the results form the
tight-binding calculation rather well up to the value $w=2 r_C$ where the
quantum Hall effect evolves. For small cyclotron radii with $w
\gtrsim r_C$ there is essentially no difference between ring and disk
geometry since the number of trajectories touching the inner
boundary gets significantly reduced.

The conductance oscillations due to the magnetic focusing are easily
distinguishable from universal conductance fluctuations.  The magnitude is
larger than a conductance quantum as multiple modes simultaneously fulfill the
focusing condition. 

The effect is only dependent on the cyclotron radius and the device geometry.
For an estimation of the thermal
stability, we compare the magnitude of the thermal fluctuations of the
cyclotron radius $\Delta r_C \approx k_B T / v_F e B$ with the width of the ring arm $w$. For the experimental parameters ($B =$1\,T and 
$T\le 4.2\,$K), we obtain the estimate $\Delta r_C \approx 0.4\,$nm which is orders
of magnitude smaller than $w$.

\subsection{Edge roughness}

After modeling the basics of magnetic focusing, we concentrate now on the extension of the simulation for a better accordance to the experimental findings.
In the experiment no quantized conductance is observed (horizontal features) and the oscillations due to magnetic focusing appear as plateaus in the magnetoconductance (compare Fig. 4(e))
Different from the experimental
situation, we have considered a perfect graphene ring without any disorder and
edge roughness so far. From the large mobility measured in the system, we
conclude that impurity scattering is negligible in the bulk of the device.
Therefore, we attribute the difference between the experimental data and the
simulation to edge roughness at the graphene boundaries. For a simple model of edge roughness, we modulate
the radius of the outer ring $r_o(\theta) =  r_o + \delta r \cos(n \theta)$
with  $n=20$ and variable amplitude $\delta r$.\cite{inner} Note that we do
not intend to realistically model the edge roughness of the experimental
situation. One of the reason is that the chemistry at the edge is not
sufficiently understood.
But even more importantly, as we simulate a
scaled version of the graphene lattice, edge defects that appear on the scale
of single atoms cannot be included even in principle. As a result, we model
the edge roughness by a single effective parameter $\delta r$ in the simple
model given above. Figure~\ref{fig:numericsb}(b) shows the effect of the edge
roughness on the magnetic focusing in the disk geometry. As it is evident from
the plot, the features due to magnetic focusing are rather robust toward the
inclusion of edge roughness.

Fig.~\ref{fig:numericsb}(c) makes the connection to the experiment as it displays
the situation of a graphene ring with the effect of edge roughness
taken into account. We see that increasing the edge roughness from $\delta
r=0$ to $\delta r= 6.3\,$nm starts suppressing the features of quantization
and magnetic focusing.  At an edge-roughness of about $\delta r=16\,$nm the
features are transformed to a more shoulder-like behavior (e.g., between the
two vertical dashed lines). The experimental situation is best described with this level of edge disorder.\cite{note1}


\bibliographystyle{apsrev4-1}

\begin{thebibliography}{99}
\bibitem{FNgeometry}
	From the SFM image we extract a 4W path length of $L \approx 3\,\mu$m and
	with accounting twice the ring width $w=350\,$nm for path width $W$ we estimate the aspect ratio $L/W\approx 4$.
\bibitem{FNmfp}
	The mean free path $l_m = \hbar \mu \sqrt{\pi \alpha |V_G-V_{CNP}|}$ is
	calculated as a lower and an upper estimate for $|V_G-V_{CNP}| = 5 - 30\,$V
	and $\mu=60{,}000-100{,}000\,$cm$^2$/Vs.
\bibitem{FNabo}
	The periodicity of AB oscillations can be determine by the sample geometry $\Delta
	B_{AB}= h/e \pi r^2$, where the radius $r$ is set to $r_{i}=405\,$nm, $\overline{r}=580\,$nm and
	$r_{o}=755\,$nm, respectively.
\bibitem{FNvisib}
	We define visibility as the ratio of AB amplitude and the 2W conductance at zero magnetic field $\Delta G_{AB}/G_{2W}(B=0)$.
\bibitem{inner} We only include edge roughness of the outer edge as the physics is dominated by this edge. We have checked that edge roughness on the inner edge has a negligible impact on the final results.
\bibitem{note1} Moreover, we observe that the shoulder-like curve even turns on a rather monotonous behavior at higher strength of the edge disorder.
\end{thebibliography}

\end{document}